\documentstyle[amssymb,aps]{revtex}
\input epsf.sty
\def\beq{\begin{equation}}
\def\eeq{\end{equation}}
\begin{document}

\title{``Weather'' Records: Musings on Cold Days after a Long Hot
Indian Summer}
\medskip
\author{B. Schmittmann and R. K. P. Zia
\bigskip}
\address{Center for Stochastic Processes in Science and
Engineering and
Department of Physics, \\
Virginia Polytechnic Institute and State University, Blacksburg,
Virginia 24061-0435}

\maketitle

\begin{abstract}
We present a simple, pedagogical introduction to the statistics of
extreme values. Motivated by a string of record high temperatures
in December 1998, we consider the distribution, averages and
lifetimes for a simplified model of such ``records.'' Our ``data''
are sequences of independent random numbers all of which are
generated from the same probability distribution. A remarkable
universality emerges: a number of results, including the lifetime
histogram, are universal, that is, independent of the underlying
distribution.
\end{abstract}

\section{Introduction and motivation}
In December 1998, in the aftermath of El Nino and its companion, La
Nina, the weather in the Roanoke, Virginia, area was unusually
mild. Weather data have been collected here since 1934, and record
highs and lows for any particular day are known. As part of the
daily weather forecasts, local TV stations report these record
values and compare them to the highest and lowest temperature
values of the day. Remarkably, during the {\em nine} days from
November 29 to December 7, 1998, the previous record highs were
{\em broken five times and tied once}!\cite{www} One might
wonder, as we did, how frequently such a series of records could
possibly occur. When only few weather data are available, such as
in the early years of record keeping, it is obviously quite easy to
experience new extremes. However, as the data sets become larger,
the record highs (and lows) are pushed to higher
(lower) values, so that the setting of a new record becomes a far
less frequent event. Thus, we began exploring questions such as:
How probable was it to set a new record in 1998, 64 years after the
records began? How do record highs increase with time? How long do
records typically survive before they are broken?

Not surprisingly, similar questions have been posed before. It
appears that the earliest studies are due to N.\ Bernoulli who
analyzed life expectancies in 1709 \cite{evs}. Later, flood
control and structural safety issues were considered, to name just
a few of the numerous applications. Beginning in the 1920's, the
mathematical techniques were developed and the study of records, or
extremes, became known as ``extreme value statistics,''\cite{evs}
which is still an active area of research. In this paper, we
provide a pedagogical introduction to some of the most basic
results. We begin in Section II with the simplest 
model for records and a concise statement of the problem. There is
no attempt to address real records. As a result of complex
physical processes, the statistics of real weather records is
undoubtedly far more intricate than that generated by simple random
numbers. The quotation marks in the title should remind the reader
that actual weather records are {\em not} the subject of this
article. Section III is devoted to the full distribution function
for the model records, their averages, and standard deviations.
Next, we discuss the record lifetimes and derive the associated
distribution in Section IV. Although we focus on ``record highs,''
a completely analogous line of reasoning can be
pursued for ``record lows.''\cite{evs} In the final section, we
turn to more general questions and conclude by listing a number of
problems which, to the best of our knowledge, are still unsolved.
Some technical details are provided in the Appendix.

\section{A simple model for records}

Our much simplified model for the physical data (temperatures, water levels,
etc.) is based on a probability density $p(x)$ for a real variable
$x$. For example, if we are considering temperatures, $p(x)dx$ might
be the probability that the temperature at noon, on a specific day
of the year, takes a value between $x$ and $x+dx$. The ``data'' in
our simplified model are just a sequence of random numbers: 
\begin{equation}
\left\{ x(1),x(2),x(3),\ldots,x(t),\ldots \right\} .
\label{sequence}
\end{equation}
In keeping with the language of weather records, we refer to the
(integer) label of these numbers as ``time'' (or ``year''):
$t=1,2,3, \ldots$ Each of the $x$'s is drawn from the {\em
same} {\em distribution} $p(x)$. Thus, our data form a set of
independent, identically distributed random variables. This
condition is probably the most serious shortcoming when applied to
physical reality, where major correlations or variations can be
expected. For example, it precludes ``global warming,'' a situation
in which the underlying distribution $p(x)$ itself is a function of
time.

Before discussing the sequence and the records, let us provide some
details about the distribution $p(x)$. It may be defined over the
entire real axis or restricted to an interval, that is, it may have
infinite or finite support. For simplicity, we require it to be
{\em continuous and positive}, which excludes, for example, dice
throwing. As a result, we can ignore the possibility of ties. Of
course, it must be properly {\em normalized}, that is, $\int
\! p(x)dx=1$. Here and in the following, any integral limits that
are not explicitly specified are to be taken as the appropriate
(upper and/or lower) bounds of the support. Because the distribution
$p(x)$ reflects how we model our data set, we refer to it as the
``model distribution,'' or simply ``the model.'' In particular, we
will investigate to what extent our results for record
distributions and lifetimes depend on the underlying model. For
later reference, we also introduce a second distribution, simply
related to $p(x)$:
\begin{equation}
q(R)\equiv \! \int^R \! p(x)\,dx \label{q-def}
\end{equation}
Clearly, $q(R)$ is the probability that a random number, drawn from the
distribution $p(x)$, will not exceed $R$. Thus, $q(R)$ is a
monotonic function, varying from 0 to 1. 
The reader familiar with random numbers on a computer will recognize 
that the inverse of this function, $R(q)$, is just the operation to 
generate random numbers for an arbitrary density $p(x)$ from the 
uniform distribution on the interval $[0,1]$. 
Following Galambos\cite{evs}, we call $q(R)$ the ``common distribution
function.''

Returning to the sequence (\ref{sequence}), we define the ``record'' $R(t)$
as the {\em largest} number in a string of length $t$: 
\begin{eqnarray}
R(t) &\equiv &\max \left\{ -\infty
,x(1),x(2),x(3),\ldots,x(t)\right\} 
\nonumber \\
&\equiv &\sup \left\{ x(1),x(2),x(3),\ldots,x(t)\right\} ,
\label{record}
\end{eqnarray}
where we have included $R(0)=-\infty $ as the (arbitrary) initial
value. After generating the next new number $x(t+1)$, we determine
whether the record has been broken, according to 
\begin{equation}
R(t+1)=\left\{ 
\begin{array}{l}
R(t)\text{ \qquad if \hspace{0.05in}}x(t+1)\leq R(t) \\ 
x(t+1)\hspace{0.05in}\hspace{0.05in}\text{if
\hspace{0.05in}}x(t+1)>R(t) .
\end{array}
\right. \label{rule}
\end{equation} 

This procedure is continued until we have obtained a sequence of length $N$.
Of course, the records $R(t)$ themselves are stochastic quantities. So, we
can define a density for $R$:
\[
P(R,t)
\]
so that the probability for finding the record to lie between $R$ and $R+dR$
is just $P(R,t)dR$. Our goal is to determine, given the underlying model
distribution $p(x)$, this $P(R,t)$ or, at a simpler level, the average
record:

\begin{equation}
\left\langle R(t)\right\rangle =\!\int \!dR\,R\,P(R,t)\,.  \label{averageR}
\end{equation}

Before delving into the analytic aspects, let us consider computer
simulations of these records. To study their statistics,
we generate a large number, $M$, of sequences (modeling, for example,
weather data from $M$ cities): $R_{i}(t),i=1,...,M$ . Based on
this ensemble of sequences, we can define the average record as a function
of time
\[
\left\langle R(t)\right\rangle =\frac{1}{M}\sum_{i=1}^{M}R_{i}(t).
\]
As an ilustration, following are \emph{two }($M=2$) sequences of \emph{ten }(%
$N=10$) random numbers generated by RAND() in MS-EXCEL (quoted to 4 digits
for simplicity):

\[
x_{1}(t)=%
\{0.6855,0.3140,0.4555,0.8082,0.0023,0.9722,0.9372,0.1412,0.7057,0.9699\}
\]

\[
x_{2}(t)=%
\{0.1042,0.8997,0.7126,0.4638,0.2598,0.9864,0.2586,0.0577,0.7777,0.8863\}
\]
and the associated sequences of records:

\[
R_{1}(t)=%
\{0.6855,0.6855,0.6855,0.8082,0.8082,0.9722,0.9722,0.9722,0.9722,0.9722\}
\]

\[
R_{2}(t)=%
\{0.1043,0.8997,0.8997,0.8997,0.8997,0.9864,0.9864,0.9864,0.9864,0.9864\}
\]
The \emph{average} record is, therefore,

\[
\left\langle R(t)\right\rangle
=\{0.3949,0.7926,0.7926,0.8540,0.8540,0.9793,0.9793,0.9793,0.9793,0.9793\}
\]
Since RAND() outputs a random number uniformly distributed  between 0 and 1,
it is hardly surpirsing to see that the average record bumps up against 1
here. To get a good grasp of this process, the reader should attempt, say,
100 sequences of $N=300$ on a spreadsheet. On a typical modern PC, it takes
only ``the blink of an eye'', literally! 

For the computer simulations presented in this article, we used a high
quality random generator 
(RAN2 from {\sl Numerical Recipies}\cite{num-rec}) and averaged over 
$M=10^{5}$ sequences of length $N=10^{3}$. This procedure requires just a 
few minutes running LINUX on a 450 MHz Pentium II, and produces excellent
statistics. In fact, with $10^{5}$ sequences, it is possible to generate a
reasonable picture of the whole probability distribution for the records. A
theoretical approach to the averages will be our next task.

\section{The record probability distribution}

\subsection{The evolution equation and its solution}

Our goal in this section is to find an analytic form for the 
probability density $P(R,t)$. 
We will do this recursively, by assuming
that we know the value of the record, say $R'$, at time
$t-1$. Then, we seek the {\em conditional} probability,
$P(R,t|R',t-1)$, that the record, at time 
$t$, has the value $R$, provided it had the value $R'$ at time
$t-1$. Fortunately, it is very easy to write down this quantity!
Clearly, it vanishes for $R<R'$, because the new record
cannot be smaller than the old one. This leaves us with two
possibilities: Either the old record stays the same so that
$R=R'$, or it is broken, resulting in the new value
$R>R'$. The former is the case if the new random number,
$x(t)$, does not exceed $R'$. Reviewing Eq. (\ref{q-def}),
we see that this case occurs with probability
$q(R)$. In contrast, if $x(t)$ exceeds $R'$, then it
sets the new record $R$. This latter case occurs with probability
$p(R)$. Summarizing, the conditional probability is

\begin{eqnarray}
P(R,t|R',t-1) &=&\left\{ 
\begin{array}{ccc}
0 & & R<R' \\ 
q(R) & \text{ for} & R=R' \\ 
p(R) & & R>R'
\end{array}
\right. \nonumber \\
&=&q(R)\delta (R-R')+p(R)\Theta (R-R')\ .
\label{conditionalP}
\end{eqnarray}
The Heavyside step function $\Theta $ i unity if its
argument is positive and zero otherwise. Its derivative is just the
Dirac delta function $\delta$, so that the two terms in
Eq. (\ref{conditionalP}) can be combined into
\begin{equation}
P(R,t|R',t-1)=\frac{\partial} {\partial R}\,\left[
q(R)\Theta (R-R')\right] \ . \label{condP}
\end{equation}
Because we have exhausted all possibilities for $R$, the conditional
probability must be normalized with respect to an integration over
$R$. This condition is easily checked: because $P$ is a total
derivative, its integral is just $q\Theta $ evaluated at the
limits. So, we have
$\int \! dR\,P(R,t\,|\,R',t-1)=1$.

From the {\em conditional} probability, it is a simple step to
arrive at our main target: the record probability density (the
``record'' distribution)
$P(R,t)$: 
\begin{equation}
P(R,t)=\!\int \! dR'P(R,t\,|\,R',t-1)P(R',t-1)\,\,\,.
\label{record_prob}
\end{equation}
Substituting the explicit form (\ref{condP}) for the conditional
probability, we obtain a recursion relation for $P$: 
\begin{eqnarray}
P(R,t) &=&\! \int \! dR'\,\frac \partial {\partial
R}\,\left[ q(R)\Theta (R-R')\right] P(R',t-1)
\nonumber \\ &=&\frac \partial {\partial R} \! \left[ q(R) \! \int
\! dR'\,\,\Theta (R-R')\right] \! P(R',t-1)
\nonumber \\ &=&\frac \partial {\partial R}q(R) \! \int^R
\! dR'\,\,P(R',t-1) .
\label{P_rec}
\end{eqnarray}
Clearly, this form is still slightly unwieldy, due to the integration. Let
us define the ``barrier'' distribution $Q(R,t)$ to be the
probability that at time $t$ the record is at $R$ or lower: 
\begin{equation}
Q(R,t)\equiv \! \int^R \! dR'\,P(R',t),
\label{Qdefined}
\end{equation}
so that 
\begin{equation}
P(R,t)=\frac \partial {\partial R}Q(R,t)\ . \label{dQdR}
\end{equation}
From Eqs. (\ref{P_rec} - \ref{dQdR}), the function $Q$ satisfies a
very simple recursion equation, namely, 
\begin{equation}
Q(R,t)=\,q(R)Q(R,t-1) . \label{Q_rec}
\end{equation}
Equation (\ref{Q_rec}) has a simple interpretation. Recall that
$q(R)$ is the probability that the next random number is less or
equal to $R$, and
$Q(R,t)$ is the probability that, at time $t$, the record has {\em
not exceeded} $R$. The product of the two gives the
probability that the record remains unbroken after the next time
step.

The recursion relation (\ref{Q_rec}) is easily solved: 
\begin{equation}
Q(R,t)=\left[ q(R)\right]^t Q(R,0)=\left[ q(R)\right]^t 
\label{Q=q^t}
\end{equation}
because $Q(R,0)=1$ for all $R>-\infty $. We can deduce two important
properties from this general solution for this barrier
distribution $Q$. Because $q(R)$ is a monotonic function, so is
$Q(R)$. This behavior of $Q(R)$ may be interpreted as at any given
time, it is more difficult to go over a higher barrier (break a
record). On the other hand, because
$q<1$ for any fixed $R<1,$ $Q$ {\em decreases} with $t$ which
implies that any given record can be broken, provided we wait long
enough! We emphasize that these ``sensible'' results are completely
independent of the details of the underlying distribution $p(x)$.

From (\ref{Q=q^t}), the record distribution follows: 
\begin{equation}
P(R,t)=\frac{\partial} {\partial R}Q(R,t)=t\left[ q(R)\right]
^{t-1}\frac{\partial q(R)}{\partial R} = t\left[ q(R)\right] .
^{t-1}p(R)
\label{P_simple}
\end{equation}
Once $P(R,t)$ is known, we can compute average values for the records, 
$\left\langle R(t)\right\rangle $, through the defintion (\ref{averageR}), as
well as the standard deviations and all higher moments. For later reference,
we provide a simplified representation of the integral in Eq.(\ref{averageR}%
). Substituting (\ref{P_simple}) into (\ref{averageR}), we have $%
\left\langle R(t)\right\rangle =\!\int \!dR\,Rt\left[ q(R)\right] ^{t-1}p(R)$%
. Next, recall that $q(R)$ is monotonic, so that it can be inverted uniquely
to give the function $R(q)$. Now, change the integration variable from $R$
to $q$, using $p(R)dR=dq$. The result is 
\begin{equation}
\left\langle R(t)\right\rangle =t\!\int_{0}^{1}\!dq\,R(q)\,q^{t-1}\ .
\label{Rtoq}
\end{equation}
Note that the limits of integration are now \emph{unique}, that is,
independent of the underlying model!

In the next section, we will illustrate the characteristics of $P(R,t)$ and
$\left\langle R(t)\right\rangle $ with two simple, explicitly
solvable distributions $p$.

\subsection{Two exactly solvable examples: flat and purely exponential
distributions}

Let us illustrate our results by considering two particularly simple cases:
a flat distribution 
\begin{equation}
p(x)=1\text{\ \ for}\ \ 0\leq x\leq 1 \label{p-flat}
\end{equation}
and a pure exponential 
\begin{equation}
p(x)=e^{-x}\text{\ \ for}\ \ 0\leq x\ . \label{p-exp}
\end{equation}
These distributions differ significantly in that the former has an
{\em upper bound} for the allowed values of the data. As a result,
the possible record values are also bounded! In contrast, the
latter distribution extends to infinity, setting no limits on the
possible records. Studying these two simple distributions will lead
us to find rather different, but hopefully generic behavior for
bounded versus unbounded models.

The flat distribution (\ref{p-flat}) corresponds to data whose
values are equally probable over a given interval. From
Eq. (\ref{Q=q^t}), we find
$q(R)=R$ (for $0\leq R\leq 1$) and
\begin{equation}
Q(R,t)=\left[ q(R)\right] ^t=R^t \label{Q-flat}
\end{equation}
so that 
\begin{equation}
P(R,t)=\partial Q/\partial R=tR^{t-1}\ . \label{P-flat}
\end{equation}
Both results are easily interpreted. The barrier distribution
$Q(R,t)$ displays explicitly the general properties discussed
above: increasing with
$R$ at fixed $t$ and decreasing with $t$ at fixed $R<1$. In
contrast, the record distribution $P(R,t)$ displays a {\em maximum
in} $t$ for a fixed $R$. For early times, the probability to
find the record at $R$ is low because this value has not yet been
reached, whereas, for late times, it has already been exceeded!

We can also study $P(R,t)$ as a function of $R$ for a fixed $t$.
The maximum value always occurs at $R=1$, increasing linearly with
time. If we apply the normalization condition $\int \! P dR=1$, the
width of the peak must narrow with time. In other words, the late
time records are ``piled up'' just below the highest allowed value
(unity in this case). Let us investigate how this feature is
reflected in the {\em average} record. Using Eq. (\ref{averageR}),
we find easily 
\begin{equation}
\left\langle R(t)\right\rangle
=t \! \int_0^1 \! dR\,R\,R^{t-1}=\frac{t}{t+1}\ .
\label{avR-flat}
\end{equation}
As expected, $\left\langle R(t)\right\rangle $ increases monotonically as a
function of time, reaching its upper bound at $t=\infty $: 
\beq
\lim_{t \to \infty} \left\langle R(t)\right\rangle =1\text{.} 
\eeq
Of course, its {\em rate} of increase must vanish in this limit. In
this case, the asymptotic is $t^{-2}$. This behavior is illustrated
in Fig. 1a, which shows that the exact result, Eq. (\ref{avR-flat})
is in excellent agreement with Monte Carlo data averaged over
$10^5$ sequences with 1000 entries each.

Several of these properties are generic in the following
sense. If the underlying distribution $p$ has an upper bound, that
is, $p(x)=0$ for
$x\geq B$, then $\lim_{t \to \infty} \left\langle
R(t)\right\rangle =B$. Not surprisingly, the behavior of $p$ near
$B$ will dictate the asymptotics. For example, if we assume
$p(x) \to \mu k\left( B-x\right) ^{\mu -1}$, so that
$q(R) \to 1-k\left( B-R\right) ^{\mu}$ for $x$, then $R \to
B$, and we can show that $B-\left\langle R(t)\right\rangle \to
\Gamma \left(1/\mu \right) \left( kt\right)^{-1/\mu}\,$, which
is a generalization of the flat case.

We next turn to the purely exponential distribution (\ref{p-exp})
and its associated $q(R)=1-p(R)$. The two characteristic
distribution functions are 
\begin{equation}
Q(R,t)= \left[1-e^{-R}\right ]^t \label{Q-exp}
\end{equation}
and 
\begin{equation}
P(R,t)=t\,e^{-R} [1-e^{-R}]^{t-1}\ . \label{P-exp}
\end{equation}
Many properties are qualitatively similar to the flat case. One difference
is that the position, $R_0(t)$, of the maximum in $P(R,t)$
increases with time indefinitely: 
\[
R_0(t)=\ln t \text{ .} 
\]
Once again, this behavior is reflected in the average record. Using
Eq. (\ref{Rtoq}) and deferring details to Appendix A, we obtain 
\begin{equation}
\left\langle R(t)\right\rangle =\sum_{k=1}^t\frac{1}{k} ,
\label{avR-exp}
\end{equation}
which is obviously a monotonically increasing function of time. To
exhibit the asymptotic behavior for large times, we can write
Eq. (\ref{avR-exp}) in more compact notation:
\beq
\left\langle R(t)\right\rangle =\psi \left( t+1\right) -\psi \left(
1 \right) \, ,
\eeq
where $\psi$ is the digamma function with known asymptotic
form.\cite{AS} As a result, 
\begin{equation}
\left\langle R(t) \right\rangle \simeq \ln t + \gamma +O(1/t) \quad
 \text{for} t \to \infty \text{ ,} \label{avR-exp-as}
\end{equation}
where $\gamma \simeq 0.5772$. Note that the {\em rate} of increase also
vanishes with time, but with a slower decay, $t^{-1}$. In Fig. 1b,
we show the comparison of the asymptotic result (\ref{avR-exp-as})
with Monte Carlo data. The agreement is clearly excellent.


\begin{figure}[tbp]

\vspace*{0.5cm}
\hspace*{-0.5cm}
\epsfxsize=7in
\epsfbox{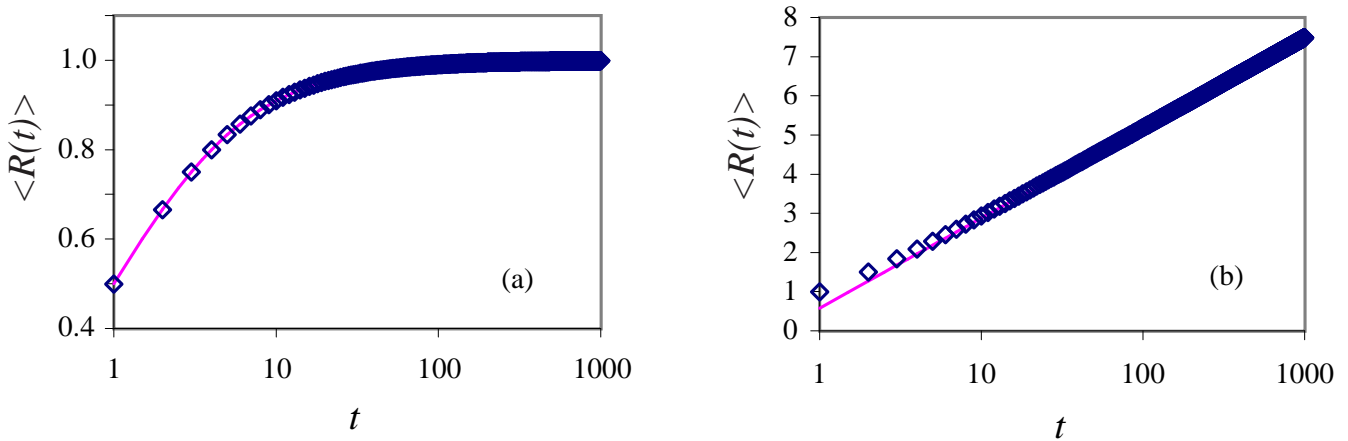}
\vspace*{0.0cm}

\caption{Average record $\left\langle R(t)\right\rangle$ versus  
time
$t$, for (a) flat and (b) exponential distributions. The diamonds
are Monte Carlo simulation data;
the solid line are the theoretical results, Eq. (\ref{avR-flat})
and Eq. (\ref{avR-exp-as}), respectively.}

\end{figure}


As before, we can consider the behavior of the average record for
more general unbounded distributions $p(x)$. Here, the argument
rests on a saddle-point approximation for $P(R,t)$ around its
maximum value,
$R_0(t)$. Asymptotically, the average can be approximated by the
position of the maximum. Because $R_0(t)$ becomes very large for
late times, the asymptotics should be controlled by the behavior of
$p(x)$ for large $x$. Some examples of possible asymptotic
behaviors of unbounded $p$'s and their associated saddle points
$R_0(t)$ are 

\begin{enumerate}
\item a power law $p\sim 1/x^\alpha $ (with
$\alpha >2$ to ensure that $\left\langle R\right\rangle $ exists),
resulting in $R_0(t)\sim t^{1/\left(
\alpha -1\right)}$;

\item an exponential $p\sim \exp (-\mu x)$, giving
$R_0(t)\sim\ln t$;

\item a Gaussian $p\sim \exp (-\mu R^2)$for which $R_0(t)\sim
\sqrt{\ln t}$.
\end{enumerate}

We invite the reader to carry out simulations for, say,
\[
p(x)=2/x^{3},\quad x\in \left[ 1,\infty \right] 
\]
and check the results against the predictions in item 1. We would
also like to caution the reader that the approach to
asymptopia for the Gaussian is {\em extremely} slow. 
In particular, we find that this regime lies beyond $t=10^{6}$.
 
\section{The distribution of record lifetimes}

Next, we turn to a key question in the study of floods or
earthquakes. What is the typical time span between two large
events? At the practical level, how much time do we have to
construct dams or to repair levees before the next record-breaking
flood? Of course, our study will not provide an {\em exact time
span} until the next disaster; it will only give an {\em estimate}
of how long a record might survive. We refer to the time span
between the setting of a record and its subsequent breaking as its
``lifetime'' (or ``record time''\cite{gal}). In the following, we
investigate the distribution of these record lifetimes.
Specifically, we will consider data sequences with $N$ entries. For
each sequence, we will identify the associated records and their
lifetimes. By analyzing a large number of data sequences, we can
compile a histogram, $T_N(m)$, of how likely a record would
survive for a time span $m$.

\subsection{A tree of lifetimes}

Let us introduce a simple representation of all possible histories of
records: a tree-like structure. We begin by generating a data
sequence
$\left\{ x(1),x(2),x(3),\ldots,x(N)\right\} $. Because we only need
to know {\em where} the records occur in this sequence, we can
associate this sequence with the following binary string. If $x(i)$
is a new record, replace it by the letter $R$ (``record'');
otherwise, replace it by $L$ (``lower''). Note that, had we used a
discrete underlying distribution $p$, then we would have to
consider the complication of ties. As an example, the sequence 
$\left\{ 0.2,0.4,0.3,0.1,0.6,0.7,0.2,0.4,0.8\right\} $ is replaced
by
$\left\{ R,R,L,L,R,R,L,L,R\right\} $. By convention, the first
entry is always $R$. If a record is established at time $t$ and
broken at time $t+m$, then that record's lifetime is defined to be
$m$. Clearly, the binary string is much simpler than the original
data sequence, but it contains enough information about record
lifetimes for us to predict the distribution
$T_N(m)$.

The binary strings are easily visualized via a tree structure. Starting from
a single vertex (the ``ancestor'') $R$ on the first line ($t=1$),
time runs downwards. At time $t=2$ (the second line), our string
has two possible entries: $R$ or $L$. To represent these, we draw
two branches from the first line to the second: one to the right
(labeled $R$), and one towards the left (labeled $L$).
Each of these branches ends in a new vertex. These branch again,
giving us four vertices in total on the third ($t=3$) line.
Continuing this procedure to the $N$th level, we find $2^{N-1}$
vertices there. As an illustration, the $N=4$ tree is shown in
Fig. 2a. Clearly, all possible binary strings with 4 elements
appear in this tree, each associated with exactly one vertex on the
$t=4$ line.

The record lifetimes associated with a given string are now easily
identified: following an $L$ branch means that the current record
survives, while choosing the $R$ branch implies that a new record
is set. Thus, each vertex can be labeled with the set of record
lifetimes $\left\{
\tau_1,\tau_2,\ldots,\tau_k\right\}$ leading to it. Fig. 2b
shows the ``tree of lifetimes'' for the $N=4$ case shown in
Fig. 2a. Note that the number of entries in these sets varies for
different strings. For example, the far right string in Fig. 2a,
where every record is broken at the next time step, gives rise to
$\left\{ 1,1,1,1\right\}$, while the far left string corresponds to
$\left\{4\right\} $: the record is set at $t=1$ and survives the
next three time steps. A simple recursion relation emerges. From a
particular entry at time $t$ to the two ``daughters'' at time
$t+1$, the set
$\left\{
\tau_1,\tau_2,\ldots ,\tau_{k-1},\tau_k \right\}$ branches into
two:
$\left\{ \tau_1,\tau_2,\ldots ,\tau_{k-1},\tau_k+1\right\}$
and $\left\{
\tau_1,\tau_2,\ldots ,\tau_k,1\right\}$. Moreover, because any
distribution $p(x)$ will generate the same set of binary strings,
it is obvious that the associated lifetimes $\left\{
\tau_1,\tau_2,\ldots ,\tau_k\right\} $ are completely {\em
independent} of the underlying model!

\subsection{The string probabilities}

To complete the construction of the histogram, we need to find the
{\em string} {\em probability}, that is, the probability that a
specific string will appear. At the $N$th level, each string is
associated with a specific vertex, which is uniquely labeled by
the set $\left\{ \tau_1,\tau_2,\ldots
,\tau_{k-1},\tau_{k}\right\} $. So, let us denote this probability
by $P_N(\tau_1,\tau_2,\ldots ,\tau_{k})$. At the top level
($t=1$), this probability is obviously trivial:
$P_1(1)=1$. For example, the year record keeping starts, any
temperature will be a ``record'' In the second ``year'' ($t=2$),
there are two vertices: $\left\{
\tau_1=2\right\}$ and $\left\{ \tau_1=1,\tau_2=1\right\}$,
corresponding to the first record surviving or being broken,
respectively. In terms of the original data sequence
$\{x(1),x(2)\}$, these possibilities are given by $x(1)>x(2)$ or
$x(1)<x(2)$. Let us focus first on the former case for which the
probability, $P_2(2)$, is just $\int \! dx_1 \! \int \!
dx_2\,p(x_1)p(x_2)\Theta \left( x_1-x_2\right)$. At first
sight, this expression seems to depend on the model distribution
$p(x)$. However, recalling Eq. (\ref{Rtoq}), we can transform the
integration variables to 
\begin{equation}
P_2(2)=\! \int_0^1 \! dq_1 \! \int_0^1 \! dq_2\, \Theta \left(
q_1-q_2\right) = \! \int_0^1 \! dq_1 \!
\int_0^{q_1} \! dq_2=\frac{1}{2}\ ,
\end{equation}
where we have exploited the monotonicity of $q(x)$ to replace
$\Theta \left( x_1-x_2\right)$ by $\Theta \left( q_1-q_2\right)$.
Not only is this integral trivial to compute, it is also {\em
manifestly independent} of the underlying model! In a similar
manner, we can compute explicitly that the probability of the
latter case, $P_2(1,1)$, is also 1/2.


\begin{figure}[tbp]

\vspace*{0.1cm}
\hspace*{2.5cm}
\epsfxsize=5in \epsfbox{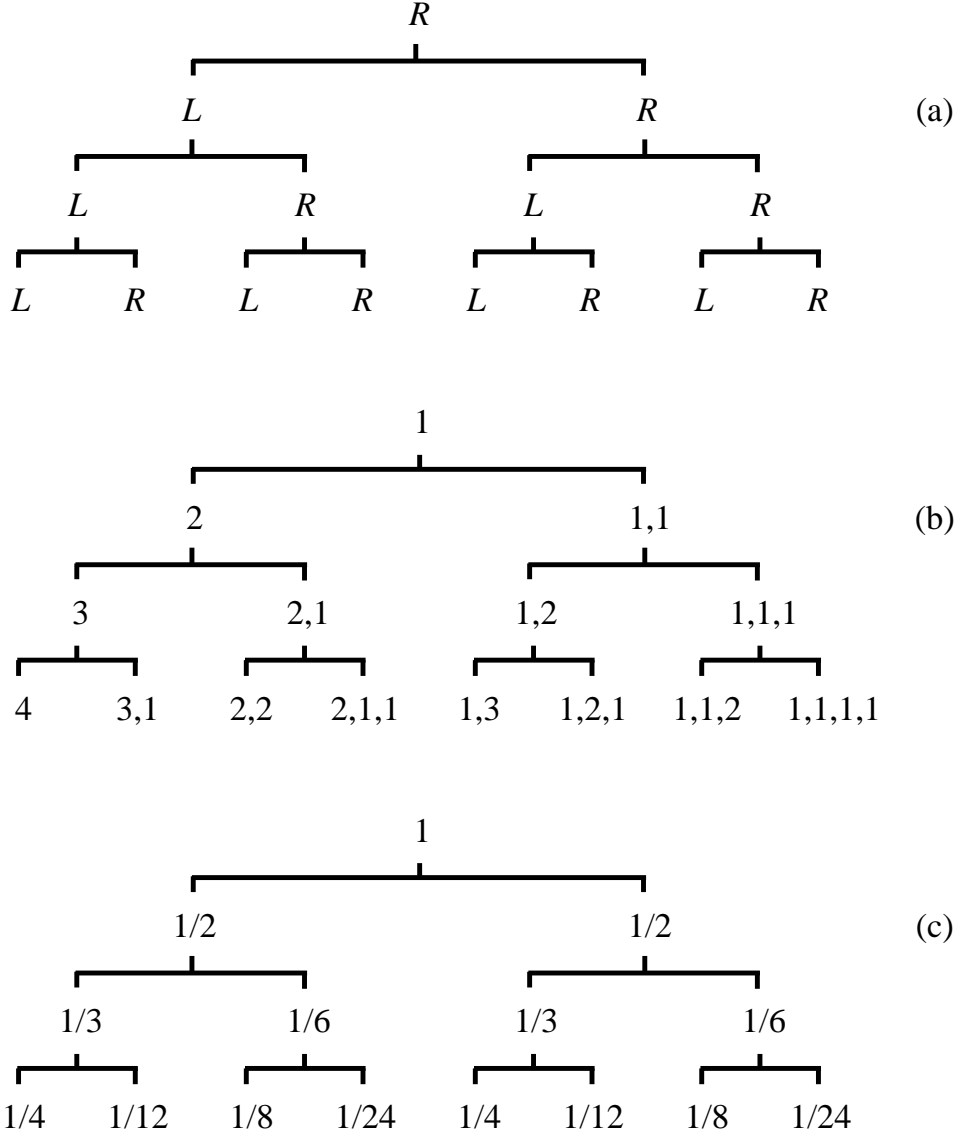}
\vspace*{0.5cm}   

\caption{Tree of records. At the $N$th level of the tree are the
$2^{N}$ possible histories. (a) The binary strings representing the
histories of records. (b) The lifetime associated with each string.
(c) Probabilities associated with each history, or ``string
probabilities."}

\end{figure}


Let us illustrate this process for the next year ($t=3$). For
the four possible histories, different combinations of
$\Theta$-functions appear. Explicitly, their associated
probabilities are:
\begin{eqnarray}
P_{3}(3)
&=&\!\int_{0}^{1}\!dq_{1}\!\int_{0}^{1}\!dq_{2}\!\int_{0}^{1}\!dq_{3}\Theta
\left( q_{1}-q_{2}\right) \Theta \left( q_{1}-q_{3}\right) =\frac{1}{3} 
\nonumber \\
P_{3}(2,1)
&=&\int_{0}^{1}\!dq_{1}\!\int_{0}^{1}\!dq_{2}\!\int_{0}^{1}\!dq_{3}\!\Theta
\left( q_{1}-q_{2}\right) \Theta \left( q_{3}-q_{1}\right) =\frac{1}{6} 
\nonumber \\
P_{3}(1,2)
&=&\!\int_{0}^{1}\!dq_{1}\!\int_{0}^{1}\!dq_{2}\!\int_{0}^{1}\!dq_{3}\!%
\Theta \left( q_{2}-q_{1}\right) \Theta \left( q_{2}-q_{3}\right) =\frac{1}{3%
}  \label{P_3} \\
P_{3}(1,1,1)
&=&\!\int_{0}^{1}\!dq_{1}\!\int_{0}^{1}\!dq_{2}\!\int_{0}^{1}\!dq_{3}\!%
\Theta \left( q_{2}-q_{1}\right) \Theta \left( q_{3}-q_{2}\right) =\frac{1}{6%
}.  \nonumber
\end{eqnarray}
In Fig. 2c, we show all the probabilities in the $N=4$
tree. Note that the sum of all the $P_N$'s at each level is indeed
unity; the probability of having {\em any} history after $N$ years
must be 1.

In Appendix B, we show that the general result for an arbitrary string 
of any length $N$ is 
\begin{equation}
P_N(\tau_1,\tau_2,\ldots ,\tau_k)=\frac {1}{\tau_1\left(
\tau_1+\tau_2\right) (\tau_1+\tau_2+\tau_3)\ldots \left(
\sum_1^k\tau_i\right)}
\, . \label{P_N}
\end{equation}
Note that the last factor is actually just $N$.

Another somewhat counter-intuitive result concerns $S_{N}(m)$, 
the probability
for the \emph{last record to survive} $m$ steps (regardless of what happened
earlier). Note that the ``last record'' is also the ``highest record,''
since the last record is necessarily larger than all previous records! Now,
to say that this record survives $m$ steps means that the values in the rest
of the string ($m-1$ of them) are lower. Therefore, in a string of $N$
steps, the last record must have occurred at the $\ell \equiv (N-m+1)$-th
step. Thus, to find $S_{N}(m)$, we ask: what is the probability for the
highest record to show up at the $\ell $-th step? We might guess that the
highest record is unlikely to occur near the beginning of the string,
because there are many chances for it to be broken later. On the other hand,
if we use the language of athletics, we might conclude that the record
facing the last athlete may be quite high (given that ``many guys have gone
before'') and breaking the record may not be easy. So, perhaps $S_{N}(m)$
should be peaked in the middle? The surprise is that $S_{N}(m)$ is \emph{%
independent of} $m$ (or equivalently, $\ell $)! In other words, the highestl
record \emph{may occur at any step with equal probability}! The proof may
be found in Appendix C.

\subsection{The universal lifetime distribution}

Lastly, we turn to the (unnormalized) lifetime distribution,
$T_N(m)$. We again llustrate by explicit calculations of
the $N=4$ case and relegate the general discussion to Appendix D.
Combining the string probabilities with the lifetimes, we obtain:
\begin{eqnarray}
T_4(4) &=&P_4(4)=\frac {1}{4} \nonumber \\
T_4(3) &=&P_4(3,1)+P_4(1,3)=\frac{1}{12}+\frac{1}{4} = \frac {1}{3} 
\nonumber \\ T_4(2)
&=&2P_4(2,2)+P_4(2,1,1)+P_4(1,2,1)+P_4(1,1,2)=\frac {1}{2} 
\label{T_4} \\ T_4(1)
&=&P_4(3,1)+2P_4(2,1,1)+P_4(1,3)+2P_4(1,2,1)+2P_4(1,1,2)+4P_4(1,1,1,1)=1.
\nonumber
\end{eqnarray}
Focusing on a particular binary string, we recall that the
associated record lifetimes $\left\{ \tau_1,\tau_2,\ldots
,\tau_k\right\}$ as well as the string probability are entirely
model-independent. As a result, the lifetime distribution itself is
{\em universal}, that is, its form does not depend on the
underlying model distribution
$p(x)$. Moreover, the result for
$T_4(m)$ suggests a surprisingly simple form for the lifetime
histogram, namely, 
\begin{equation}
T_N(m)=1/m\text{ .} \label{T_N}
\end{equation}
The general proof can be found in Appendix D. Remarkably, $T_N$ is not
only universal, but also is independent of $N$, that is, the length
of the original data sets! So, the probability for a record to
survive, for example, 5 time steps is the same, no matter how much
data we accumulate. Phrased differently, this result is not
surprising at all: because $T_N(m)$ is also, roughly speaking, the
probability for the next huge disaster to strike after 
$m$ time steps, it would be very disturbing if that probability
depended on the length of our data sets: that would imply that we
could avoid or court disaster by just continuing to take data.
Clearly, such a result would be nonsensical. Nobody would
believe that, by observing the weather, we can change it!

Although the lack of an $N$-dependence can be understood in this
way, the universality with respect to the underlying model $p(x)$
is a much stronger statement: As long as {\em all} underlying data
arise from the {\em same} distribution, the functional{\em \ form}
of this distribution is completely irrelevant. Flat, exponential or
Gaussian $p$'s all generate identical lifetime histograms. In
Fig. 3, we show Monte Carlo data for the lifetime histograms of
flat and exponential $p(x)$. They are indistinguishable, apart from
statistical fluctuations! The agreement with the theoretically
expected $T_N(m)=1/m$ is again excellent.


\begin{figure}[tbp]

\hspace*{-1.5cm}
\hspace*{0.5cm}
\epsfxsize=7in \epsfbox{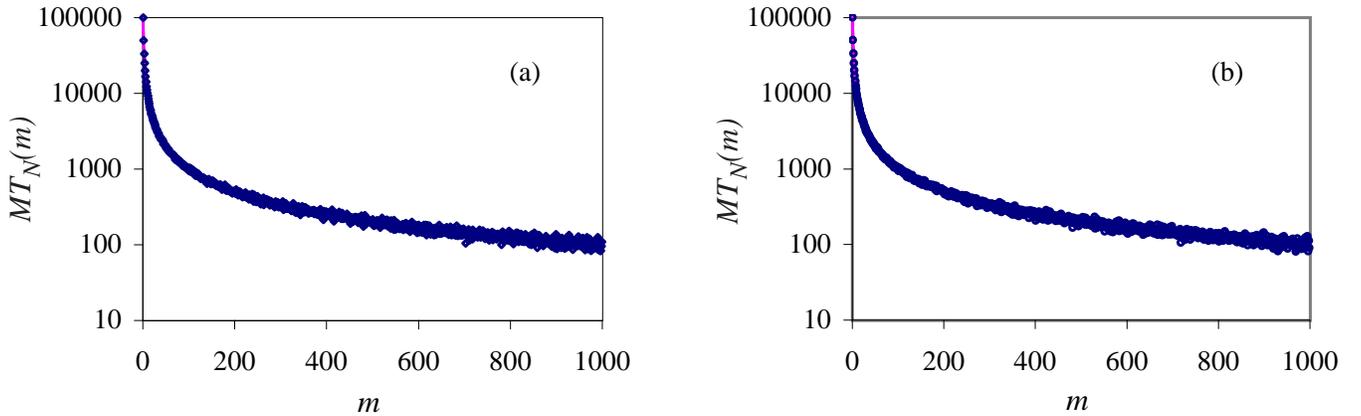}

\caption{Histogram of record lifetimes, for (a) flat and
(b) exponential distributions. The normalization factor $M$ is the
number of sequences,
$10^{5}$, that have been averaged. The theoretical line is barely  
visible behind the data points.}

\end{figure}


\section{Concluding Remarks}

We have presented a pedagogical introduction to the statistics of
extremes. Though we motivated the discussion by reference to weather
records, the model studied here is much simplified. Generating a
sequence of $N$ random numbers, all selected from the {\em same}
underlying distribution
$p(x)$, we keep track of the ones which are, say, higher than all
the preceding numbers -- ``record highs.'' Studying the statistics
of such strings, we identify several universal features, that is,
properties which are independent of the underlying distribution
$p$. In particular, we ask how long a record survives before being
broken, that is, we focus on the lifetimes of records and compile a
histogram. Not only is that histogram universal, but it is also
exceedingly simple: records lasting $m$ steps occur with a
decreasing frequency of $1/m$. We also inquire into some details of
these strings. As far as records are concerned, the sequence of
numbers can be reduced to a binary string of $R$'s and $L$'s. At
each step, we only keep track of whether the record is broken
$\left( R\right)$ or not 
$\left( L\right)$. A ``tree'' representing all possible sequences
can be drawn which displays whether a record is broken or not at
each step. It is natural to ask, What is the probability that
records will be broken at specific steps along the string?
The answer, given by Eq. (\ref{P(r)}) or (\ref{P_N}), is also
universal. Perhaps most surprising is the answer to the question,
What is the probability that the overall record will occur at the
$m$th step? The answer is {\em independent} of
$m$. The overall record occurs at any step with {\em equal
probability}.

Another interesting question is the following. Averaged over many
sequences, how does the overall record ``inch up'' with time?
Although the answer is not completely universal, the average
behavior falls into one of several classes. The simplest is due to
a $p(x)$ which is bounded. Then, not surprisingly, the average just
runs into this bound. For unbounded $p$'s, the asymptotic behavior
is mostly dictated by the ``tail'' of the distribution. In this
sense, there is a limited form of universality, that is, properties
are independent of the details of the rest of $p(x)$. Beyond
the average record, we also studied the probability density,
$P(R,t)$, for finding the record at $R$ after $t$ steps. Like the
central limit theorem, there is universality for large values
of $t$, provided an appropriate time-dependent rescaling of
$R-\left\langle R \right\rangle$ is included. However, there are
{\em three} limiting distributions, as well as the possibility of
{\em no} limiting distribution. These fascinating properties lie
outside the scope of this article. We refer the interested reader
to the book by Galambos for example.\cite{evs}

Looking beyond our simple model for ``weather'' records, we may
inquire about a natural generalization -- a model for ``global
warming.'' Here, we let the underlying distribution {\em drift
upwards} with time. The simplest is a uniform drift, that is,
letting the model distribution at time step $t$,
$p_t(x)$, assume the form $p(x-\alpha t)$ with $\alpha >0$. Although
numerous results still hold, there are also significant
differences. Obviously, the average record here must increase
linearly. Probably the most significant difference between the
simple model for ``weather'' records and this ``global warming''
case is that the limiting distribution for $P(R,t)$ is expected to
assume the form $P^*(R-\alpha t)$, with no time-dependent
rescaling. We are not aware of any general results for such 
drifting $p$'s. Certainly, these expectations are confirmed for a
{\em flat, drifting} $p_t(x)$. For large times, we find
$P(R,t)=\partial_{R}Q(R,t)$, with 
\beq
Q(R,t)=Q^{*}(R-\alpha t),
\eeq
where 
\beq
Q^{*}(\xi )=\alpha ^{k}\frac{\Gamma \left( \xi /\alpha
+1\right)} {\Gamma
\left( \xi /\alpha +1-k\right)} \quad \text{for} 1-k\alpha \leq
\xi
\leq 1-(k-1)\alpha \, . 
\eeq
To arrive at these results, we relied on a generalized version of
(\ref{Q=q^t}): $Q(R,t)=\prod_{n=1}^tq_{n}(R)$, where
$q_{n}(R)\equiv \! \int^{R} \! p_{n}(x) dx$.  Clearly, 
these general forms are applicable for {\em any} time-dependent
$p$. For example, we could consider a sequence composed of the {\em
sum} of all random numbers generated before. If the random numbers
are distributed such that both positive and negative values are
present, we may think of the sequence as the value of a stock,
making gains and losses from day to day. The variations are limited
only by the imagination. Unlike the simple model presented above,
there are very few known results for the distributions of the record
lifetimes in general. Needless to say, exploring the universality
classes will be a task both daunting and rewarding.

\begin{center}
{\bf ACKNOWLEDGMENTS}
\end{center}

We are grateful for discussions with S. Redner. This research is
supported in part by grants from the National Science Foundation
through the Division of Materials Research.

\vspace*{2.0cm}
\begin{center}
{\bf APPENDICES}
\end{center}

\subsection{The integral for (\ref{avR-exp})}

To obtain the desired result (\ref{avR-exp}), we start from (\ref{Rtoq}) and
the inverse of (\ref{Q-exp}).
\begin{eqnarray}
t \! \int_0^1 \! dq\left[ -\ln (1-q)\,\right] q^{t-1} &=& t \! \int
\!
\left[
\sum_{n=1}^\infty q^n/n\,\right] q^{t-1} \nonumber \\
&=&\sum_{n=1}^\infty \frac {t}{n\left(
n+t\right)}=\sum_{n=1}^\infty \left[ 
\frac {1}{n}-\frac {1}{n+t}\right] \label{avR-exp-ap} \\
&=&\sum_{k=1}^t\frac {1}{k} \nonumber
\end{eqnarray}
Those who like a bit more mathematical rigor may consider
$\lim_{s\to 1}\int_0^sdq \ldots$, which justifies the exchange of
the integral and sum.

\subsection{General string probabilities}

We provide some details for computing the general case
$P_N(\tau_1,\tau_2,\ldots ,\tau_k)$. First, let us introduce an
alternate way of labelling the vertices. Instead of the record
lifetimes $\left\{
\tau_1,\tau_2,\ldots ,\tau_k\right\}$, we keep track of the
times $\left\{ r_1,r_2,\ldots ,r_k\right\}$
when records are broken: that is, $r_i$ denotes the position of the
$i$th $R$ along the string. So, we may also denote the general
string probability  
by $P_N\left( r_1,r_2,\ldots
r_k\right)$. By convention $r_1=1$, while
$r_2=r_1+\tau_1$, $r_3=r_2+\tau_2$, etc: 
\begin{eqnarray*}
\tau_1 &=&r_2-r_1=r_2-1 \\
\tau_2 &=&r_3-r_2 \\
&&\vdots \\
\tau_{k-1} &=&r_k-r_{k-1} \\
\tau_k &=&N+1-r_k,
\end{eqnarray*}
where the last line comes from the length of the string:
$N=r_k+\tau_k-1$.

To obtain $P$ for an arbitrary string, we start with 
\beq
P_N\left( r_1,r_2,\ldots r_k\right) =\int dq_1\ldots dq_N\times
\Theta
\ldots \Theta \ . 
\eeq
To say that the first record $R_1$ associated with $q_1$ has a
lifetime of $\tau_1$ means that the next $\left( \tau_1-1\right)$
$R$'s (and their associated $q$'s) are lower. Therefore, the first
$\left( \tau_1-1\right)$ factors in the integrand are 
$\Theta \left( q_1-q_2\right) \Theta \left(
q_1-q_3\right)\ldots\Theta \left( q_1-q_{r_2-1}\right) 
$
so that the integration over the variables 
$q_2,q_3,\ldots ,q_{r_2-1}$
can be performed trivially, with the result $q_1^{r_2-2}$.

Now, the next record is $R_{r_2}$ associated with $q_{r_2}$, so
that the next factor in the integrand must be $\Theta \left(
q_{r_2}-q_1\right)$. There will be no more appearances of $q_1$ in
the rest of the integrand. So, the integral over $q_1$ (up to
$q_{r_2}$) can be performed, giving $
q_{r_2}^{r_2-1}/\left( r_2-1\right)$.
Note that $\tau_1=r_2-r_1=r_2-1$ so that this result can be
written simply as $
q_{r_2}^{r_2-1}/\tau_1$.

In a similar way, we integrate over the $r_3-r_2-1$ variables
$q_{r_2+1},\ldots ,q_{r_3-1}$ up to $q_{r_2}$, arriving at 
$
q_{r_2}^{r_3-2}/\tau_1$ or $q_{r_2}^{r_3-2}/\left(
r_2-1\right)$.
Performing the integral over $q_{r_2}$ gives $
q_{r_3}^{r_3-1}/\left( r_2-1\right) \left( r_3-1\right)$. This
process can be carried on until the last integration (over
$q_{r_k}$, up to 1). The integrand here consists of the result from
the integrals over the previous variables
$
q_{r_k}^{r_k-1}/\left( r_2-1\right) \left( r_3-1\right) \ldots \left(
r_k-1\right)$ as well as the $q$'s from the rest of the sequence:
$q_{r_k+1},\ldots ,q_N$. With this additional factor of
$q_{r_k}^{N-r_k}$, the last integral provides a factor of $N$. The
final result is 
\begin{equation}
P_N\left( r_1,r_2,\ldots r_k\right) =\frac {1}{\left( r_2-1\right)
\left( r_3-1\right) \ldots \left( r_k-1\right) N} \, . 
\label{P(r)}
\end{equation}
Rewriting the $r$'s in terms of the $\tau $'s, we have 
\beq
P_N\left( \left\{ \tau \right\} \right) =\frac {1}{\tau _1\left(
\tau_1+\tau_2\right) \ldots \left( \sum_1^k\tau_i\right)} ,
\eeq
where the last factor is precisely $N$, because the sum of the
lifetimes of all records is just the length of the string.

One interesting property of these $P$'s follows from the fact that,
given a particular string of length $N$, we can ``generate''
two strings of length $N+1$, by concatenating either an $R$ or an
$L$ at the end. In the former case,
$\tau_{k}$ stays the same while $\tau_{k+1}=1$. In the latter case,
the value of
$\tau_{k}$ increases by 1. From Eq. (\ref{P_N}), we obtain
the following ``recursion'' relations: 
\begin{eqnarray}
P_{N+1}\left( \tau_1,\tau_2,\ldots ,\tau_{k}+1\right)
&=&\frac{N}{N+1} P_N\left( \tau_1,\tau_2,\ldots ,\tau_{k}\right) .
\label{recur1} \\
P_{N+1}\left( \tau_1,\tau_2,\ldots ,\tau_{k},\tau_{k+1}=1\right) &=&
\frac {1}{N+1}P_N\left( \tau_1,\tau_2,\ldots ,\tau_{k}\right) .
\label{recur2}
\end{eqnarray}
Not surprisingly, 
\begin{equation}
P_{N+1}\left( \tau_1,\tau_2,\ldots ,\tau_{k},1\right)
+P_{N+1}\left(
\tau_1,\tau_2,\ldots ,\tau_{k}+1\right) =P_N\left(
\tau_1,\tau_2,\ldots ,\tau_{k}\right) \label{sum} .
\end{equation}

To illustrate, consider the first pair of entries at the 4$^{th}$ level
in Fig. 2 and their relationship to the first entry at the 3$^{rd}$ level ($%
N=3$). From Fig 2c, we read off $P_{3}\left( 3\right) =1/3$, $P_{4}\left(
4\right) =1/4$, and $P_{4}\left( 3,1\right) =1/12$. These quantities satisfy
Eqs. (\ref{recur1},\ref{recur2}). 

\subsection{Probability for strings where the last record survived
$m$ steps}

An easy way to arrive at the lifetime histogram $T_N(m)=1/m$ is
through the probability for the last record to survive $m$
steps, regardless of what happened earlier. Let us define
this quantity as $S_N(m) $. To be precise, it is 
\begin{equation}
S_N(m)\equiv \sum_{\left\{ \tau \right\}} \delta \left(
\tau_{k}-m\right) P_N\left( \left\{ \tau \right\} \right) ,
\label{Sdef}
\end{equation}
where $\delta$ is the Kronecker delta ($\delta$ is unity if its
argument vanishes and zero otherwise). Be careful with this sum
notation: it stands for summing over $k$ as well, because $k$ is
the {\em number of records} in the string. Now, because the
``last'' record is also the overall record, (that is, all other
$R$'s are lower), it is easy to compute $S_N(m)$. 
Because we are summing over all possible ways that 
the records before this step ($\ell =N-m+1$) are broken, we may write 
\begin{eqnarray}
S_{N}(m) &=&\int dq_{1}\ldots dq_{N}\times \Theta \left( q_{\ell
}-q_{1}\right) \ldots \Theta \left( q_{\ell }-q_{\ell -1}\right) \Theta
\left( q_{\ell }-q_{\ell +1}\right) \ldots \Theta \left( q_{\ell
}-q_{N}\right)   \nonumber \\
&=&\int dq_{\ell }\left( q_{\ell }\right) ^{N-1}=1/N.  \label{S=1/N}
\end{eqnarray}

As remarked in the main text, this result is not intuitively obvious
at all. Intuition might lead us to an $S_N(m)$ with a
maximum in the middle of the string, but Eq. (\ref{S=1/N}) tells us
that the distribution is entirely flat.

\subsection{Histogram for record lifetimes $T_N(m)$}

Note that $T_N(m)$ is not a real probability in the sense that
$\sum_{m}T\geq 1$. The normalization condition here is 
\beq
\sum_{\left\{ \tau \right\}} P_N\left( \left\{ \tau \right\} \right)
=1 ,
\eeq
but the histogram is defined by 
\beq
T_N(m)=\sum_{\left\{ \tau \right\}} P_N\left( \left\{ \tau
\right\}
\right) \sum_{i=1}^{k}\delta \left( \tau_{i}-m\right) \, . 
\eeq
Again, we caution the reader that the $\sum_{\left\{ \tau
\right\}}$ here involves a sum over $k$ as well. Summing over $m$
produces $\sum_{\left\{
\tau \right\}} P_Nk\geq \sum_{\left\{ \tau \right\}} P_N=1$.

Now, given a string of $N$ random numbers, the only way for a record
lifetime to be $N$ is that the first record survives. Thus,
$T_N(N)$ is precisely $P_N(\tau_1=N)$, so that 
\begin{equation}
T_N(N) = 1/N \, . 
\label{TNN}
\end{equation}
Our goal reduces to proving 
\beq
T_{N+1}(m)=T_N(m)\quad \text{for}\quad 1\leq m<N\, . 
\eeq
There are two types of contributions to $T_N(m)$. One is from all
the ``interior'' records: 
\begin{equation}
\tilde T_N(m)=\sum_{\left\{ \tau \right\}} P_N\left( \left\{
\tau
\right\} \right) \sum_{i=1}^{k-1}\delta \left( \tau_{i}-m\right) 
\label{Ttilde}
\end{equation}
that is, $\tau_{i}=m$ with $i<k$. The other piece, from the last
record alone, is precisely $S_N(m)$. For $\tilde T_{N+1}$, let
us start with $\tilde T_N(m)$ and go from $N$ to $N+1$, by
looking at 
\begin{equation}
\sum_{\left\{ \tau \right\} }P_{N+1}\left( \left\{ \tau \right\} \right)
\sum_{i=1}^{k-1}\delta \left( \tau _{i}-m\right) ,  \label{SPd}
\end{equation}
where the $k$'s appearing here are still those associated with
$\tilde T_N(m)$ (that is, the last record may be labeled by
$k+1$). Because the sum over $\tau_{k}$ and $\tau_{k+1}$ can be
performed without the $\delta $'s, Eqs. (\ref{recur1})
and (\ref{recur2}) can be used to obtain 
the following recursion relation
\beq
\sum_{\left\{ \tau \right\}} P_{N+1}\left( \left\{ \tau \right\}
\right)
\sum_{i=1}^{k-1}\delta \left( \tau_{i}-m\right) =\sum_{\left\{
\tau
\right\}} P_N\left( \left\{ \tau \right\} \right)
\sum_{i=1}^{k-1}\delta
\left( \tau_{i}-m\right) .
\eeq
But the quantity (\ref{SPd}) is not the only
contribution to
$\tilde T_{N+1}$, because concatenating an $R$ to an
$N$-string will produce an ``interior'' record for the $\left(
N+1\right)$-string. Focusing on a specific $m$, this extra bit is
just 
\beq
\sum_{\left\{ \tau \right\}} P_{N+1}\left( \tau_1,\tau_2,\ldots
,\tau_{k},\tau_{k+1}=1\right) \delta \left(\tau_{k}-m\right) ,
\eeq
which is 
\beq
\sum_{\left\{\tau \right\}} \frac{1}{N+1}P_N\left(
\tau_1,\tau_2,\ldots ,\tau_{k}\right) \delta \left(
\tau_{k}-m\right) =\frac{1}{N+1} S_N(m) .
\eeq
So, we conclude 
\beq
\tilde T_{N+1}(m)=\tilde T_N(m)+\frac{S_N(m)\text{}} {N+1} .
\eeq
With the help of this equation, we have
\begin{eqnarray*}
T_{N+1}(m) &=&\tilde T_{N+1}(m)+S_{N+1}(m) \\
&=&\tilde T_N(m)+\frac{S_N(m)\text{}} {N+1}+S_{N+1}(m) .
\end{eqnarray*}
But, according to Eq. (\ref{S=1/N}), 
\beq
S_{N+1}(m)=\frac{N\text{}} {N+1}S_N(m) 
\eeq
so that 
\beq
\tilde T_{N+1}(m)=\tilde T_N(m)+S_N(m)=T_N(m) \, . 
\eeq
Thus, using (\ref{TNN}), we arrive at 
\begin{equation}
T_N(m)=1/m\text{} \label{T=1/m}
\end{equation}
for any $N\geq m$.


\bigskip\begin{references}
\bibitem{www} The data can be found at:
http://www.wdbj7.com/climate/climate.htm.

\bibitem{evs} See for example, E. J. Gumbel, {\sl The Statistics
of Extremes} (Columbia University Press, New York 1958); J.
Galambos, {\sl The Asymptotic Theory of Extreme Order
Statistics} (Wiley, New York 1978). Gumbel gives a short historical
summary in his book.

\bibitem{num-rec} W. H. Press, S. A. Teukolsky, W. T. Vetterling,
and B. P. Flannery, {\sl Numerical Recipies, Second Edition} 
(Cambridge University Press 1992). 

\bibitem{AS} M. Abramowitz and I. A. Stegun, {\sl Handbook of
Mathematical Functions}
(Dover Publications, New York 1974).

\bibitem{gal} Record times are discussed, for example,
by Galambos (see Ref. 2).
\end{references}
\end{document}